\documentclass{PoS}
\usepackage{amsmath}

\title{Twisted reduction in large N QCD with adjoint Wilson fermions}

\ShortTitle{Twisted reduction in large N QCD with adjoint Wilson fermions}

\author{Antonio Gonz\'alez-Arroyo$^{ab}$ \\
\llap{$^a$}Instituto de F\'{\i}sica Te\'orica UAM/CSIC, C/ Nicol\'as Cabrera 13-15\\
       Universidad Aut\'onoma de Madrid, E-28048--Madrid, Spain \\
\llap{$^b$}Departamento de F\'{\i}sica Te\'orica, C-15 \\
       Universidad Aut\'onoma de Madrid, E-28049--Madrid, Spain\\
E-mail: \email{antonio.gonzalez-arroyo@uam.es}}       
\author{\speaker{Masanori Okawa}$^c$ \\      
\llap{$^c$}Graduate School of Science, Hiroshima University\\
Higashi-Hiroshima, Hiroshima 739-8526, Japan\\
E-mail: \email{okawa@sci.hiroshima-u.ac.jp}}

\abstract{The twisted space-time reduced model of large $N$ QCD with 
various flavours of adjoint Wilson fermions is constructed applying 
the symmetric twist boundary conditions with flux $k$. The models 
with one and two flavours show distinctive behaviours. 
For the two flavor case,  the string tension, calculated at $N=289$,
approaches zero as we decrease the quark mass in a way consistent with 
the theory being governed by an infrared fixed point.  In contrast,
the string tension for the case of a single  adjoint Wilson fermion remains
finite as the quark mass decreases to zero, supporting that
this is a confining theory.}

\FullConference{31st International Symposium on Lattice Field Theory LATTICE 2013\\
		 July 29 - August 3, 2013\\
		 Mainz, Germany}

\begin{document}

\section{Introduction}

The dynamics of SU($N$) gauge theories with adjoint fermions is
expected to depend crucially  on  the number of flavors $N_f$.  This
is suggested  by inspecting the $N_f$ dependence of the beta  function
$\beta$. The first two coefficients of $\beta$  expressed in terms of 't Hooft coupling
$\lambda=g^2 N$ are $b_0=(4N_f-11)/24\pi^2$ and $b_1=(16N_f-17)/192\pi^4$.  
Asymptotic freedom requires that $b_0$ should be negative, so only $N_f$=1 and 2 are allowed
(in this talk we do not  consider half-integer $N_f$ corresponding to Majorana fermions).  
For $N_f$=1, $b_1$ is also negative, so we naturally expect that the theory is confining
as ordinary QCD.  For $N_f$=2, on the contrary,
$b_1$ is positive, indicating that there could be an infrared fixed point at finite value of 
the 't Hooft coupling where the beta function becomes zero.  Since no
dimensional scale   exists at the infrared  fixed point, this theory is conjectured to be 
conformal.  In fact, for $N$=2 (minimal walking technicolor),
there are now many lattice simulations indicating that the theory is conformal at vanishing 
fermion mass\cite{DD}.

The purpose of the present talk is to study both $N_f$=1 and 2 theories in the large $N$ limit. 
It is quite obvious that the direct application of the usual lattice simulation is 
unpractical for large $N$.   Our idea is to use the twisted space-time 
reduced model defined on a $1^4$ lattice, recently proposed by the present authors\cite{GAO}.   
We point out that, in recent  years, many authors have studied space-time reduced models of 
large $N$ QCD with adjoint fermions using periodic boundary conditions\cite{KUY,AHUY,BS,HN}.
It turns out, however, that these models have too large finite $N$ corrections compared with 
those based on  twisted boundary conditions\cite{GAO,AHUY}, casting
doubts on whether the former models are of practical use. The main
finite $N$ corrections of  the twisted reduced model for $N=L^2$
amount to the finite volume corrections of ordinary Lattice Gauge
Theory on an $L^4$ lattice\cite{GAO,TEK}. Thus, by choosing
$N=17^2=289$, we can study  large $N$ QCD  with adjoint fermions on 
an effectively $17^4$ lattice within the present computer resources.
From a practical point of view, the  most important property of the reduced model is its 
rather small  memory size.  In fact, the size of four SU(289) matrices is only 5 MB, 
which can be fitted into cache memory, resulting in a rather high performance of computations.
By making use of these advantages of the reduced model, we will analyze the properties of 
large $N$ QCD with adjoint fermions and clarify the difference of $N_f$=1 and 2 fermions.  

\vspace{-0.2cm}

\section{Formulation} 

We consider the SU($N$) group with $N=L^2$, $L$ being some positive integer.  Then the action of the  
twisted space-time reduced model of QCD with $N_f$ adjoint fermions is given by\cite{GAO}

\vspace{-0.5cm}
\begin{eqnarray}
 \nonumber
   S&=&-bN \sum_{\mu \ne \nu =1}^4 {\rm Tr} \left[ z_{\mu\nu} U_\mu
  U_\nu U_\mu^\dagger U_\nu^\dagger \right]  \\
     &&- \sum_{j =1}^{N_f} {\rm Tr}\left[ \right. {\bar \Psi}^j \Psi^j  
            -\kappa \sum_{\mu=1}^4 \left\{ {\bar \Psi}^j
	    (1-\gamma_\mu) U_{\mu} \Psi^j U_{\mu}^\dagger
	     +{\bar \Psi}^j (1+\gamma_\mu) U_{\mu}^\dagger \Psi^j U_{\mu}
	         \right\} \left. \right] \label{STR2} \nonumber\\
		 &\equiv&-b N \sum_{\mu \ne \nu =1}^4 {\rm Tr} \left[
		 z_{\mu\nu} U_\mu U_\nu U_\mu^\dagger U_\nu^\dagger \right]
		 - 2 \kappa \sum_{j =1}^{N_f} {\rm Tr}\left[ {\bar \Psi}^j D_W
		 \Psi^j  \right] .
 \label{S}
 \end{eqnarray}

\noindent
$U_{\mu}$ are four SU($N$) link variables and $\Psi^j$ are $N_f$ Grassman-valued $N\times N$ matrices  
transforming in the ($N,{\bar N}$) color representation.  Spinor indices of $\Psi^j$ are not explicitly shown.  
$b$ is the inverse (lattice) 't Hooft  coupling $b=1/g^2N$ and $\kappa$ is the
hopping parameter of Wilson fermions. The symmetric twist tensor
$z_{\mu\nu}$ is an element of Z($L$), whose   explicit form is 

\vspace{-0.2cm}
\begin{equation}
 z_{\mu\nu} = \exp \left( k {2\pi i \over L} \right), \ \ \ z_{\nu\mu}=z_{\mu\nu}^*, \ \ \ \mu>\nu 
\label{Z}
\end{equation}

\noindent
The integer $k$ represents the flux through each plane.  $k$ and $L$ should be
co-prime, and a general prescription for choosing $k$ and $L$  to  minimize the 
finite $N$ corrections is given in Ref. \cite{GAO}.   
The condition is essentially the same as    the one imposed in the pure gauge model to 
prevent Z($L$) symmetry breaking\cite{TEK2},
which is necessary for reduction  to work\cite{EK,EKB}.
We recall that our prescription is to take both  $k/L$ and $\bar{k}/L$ 
(defined $k \bar{k} =1$ mod $L$) large enough.
Throughout this paper we use $L$=17 ($N=L^2=289$), $k$=5, and thus $\bar{k}$=7.
We have studied the model with $N_f=2$ by means of the Hybrid Monte Carlo method.  
For $N_f=1$, we have used  the Rational Hybrid Monte Carlo method.   

Simulations have been done at two values of the inverse 't Hooft coupling $b$ = 0.35 and 0.36.  For $N_f$=2,
we have made simulations at eight values of $\kappa$ = 0.05, 0.10, 0.11, 0.12, 0.13, 0.14, 0.15 and 0.16.
For $N_f$=1, we attempted to make simulations at the same eight values
of $\kappa$.   However, we found that, for $\kappa >$ 0.155,
the CG iteration during the molecular dynamics evolution does not
converge.  Hence,  for $N_f$=1 we took $\kappa$ = 0.05, 0.10, 0.11, 0.12, 0.13, 0.14, 0.15 and 0.155 
instead.

For every configuration we  calculated the expectation value of  ${\rm Tr}(U_\mu^\ell$), 
for $1\le \ell \le  (L-1)$,
which  are the  order parameters of the Z$^4(L)$ symmetry. We confirm
that,   in all the simulations presented here, the quantities  $<{\rm Tr} (U_\mu^\ell)>$ are compatible with zero within statistical errors.
For randomly chosen gauge configurations, we also calculated all
traces of open loops within the effective $L^4$ box,
checking that traces of all open loops are zero within statistical errors.  

\vspace{-0.5cm}
\begin{figure}[htb]
\begin{center}
\includegraphics[width=0.6\textwidth]{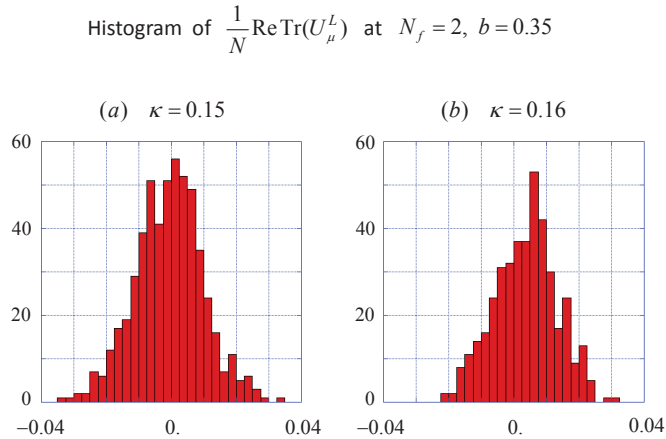}
\end{center}
\vspace{-1.cm}
\caption{Histogram of ${1 \over N} {\rm ReTr}(U_\mu^L)$ for $N_f$=2 and $b=0.35$.
(a) $\kappa$=0.15. (b) $\kappa$=0.16.}
\label{fig_histogram}
\end{figure}
\vspace{-0.2cm} 

During the simulations we also calculated ${\rm Tr}(U_\mu^L)$.  We
point out  that this quantity could have a  non-zero value without breaking 
the Z$^4(L)$ symmetry of the reduced model. The association of our
system with an ordinary lattice system of size $L^4$, indeed suggests 
that at sufficiently weak coupling a non-zero expectation value would
be observed.  However, we confirmed that $<{\rm Tr} (U_\mu^L)>$ is
statistically compatible with zero for all our simulations except for
two runs at $N_f=2$ and $\kappa=0.16$.   This is illustrated in
Fig.\ref{fig_histogram}, where  we display  the histogram of 
${1 \over N} {\rm ReTr}(U_\mu^L)$ 
at $\kappa$=0.15 and 0.16 for $b=0.35$.
While the histogram is centered at ${1 \over N} {\rm ReTr}(U_\mu^L)$=0
for $\kappa$=0.15, it is slightly shifted towards 
positive values for $\kappa$=0.16.  We observe the same phenomena at $b$=0.36.    
We expect the change of pattern to take place when a certain
correlation length of the system becomes comparable to the effective
size of the box $L$. Indeed, as shown in sect. 4, for $\kappa$=0.16 
the dimensionless  ratio   $1/(L \sqrt{\sigma})$ reaches $\sim$0.6. 

\vspace{-0.5cm}

\section{Quark mass $m_q$}

\vspace{-0.2cm}

\begin{figure}[htbp]
 \begin{minipage}{0.5\hsize}
  \begin{center}
   \includegraphics[width=70mm]{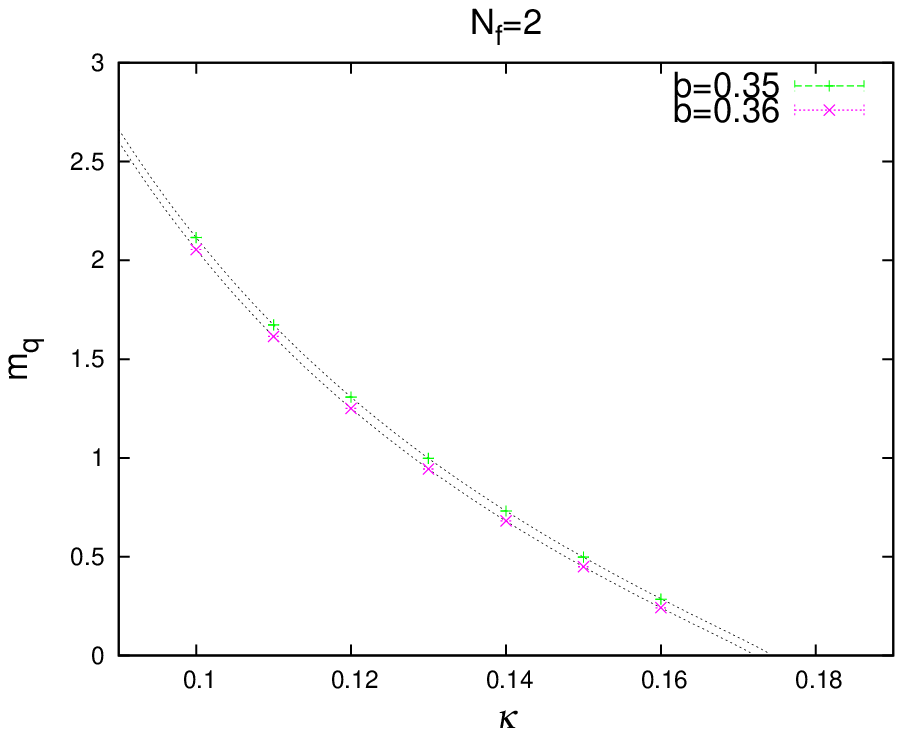}
  \end{center}
\vspace{-0.2cm}  
  \caption{$\kappa$ dependence of $m_q$ for $N_f$=2.}
  \label{fig_mq_2}
 \end{minipage}
 \begin{minipage}{0.5\hsize}
  \begin{center}
   \includegraphics[width=70mm]{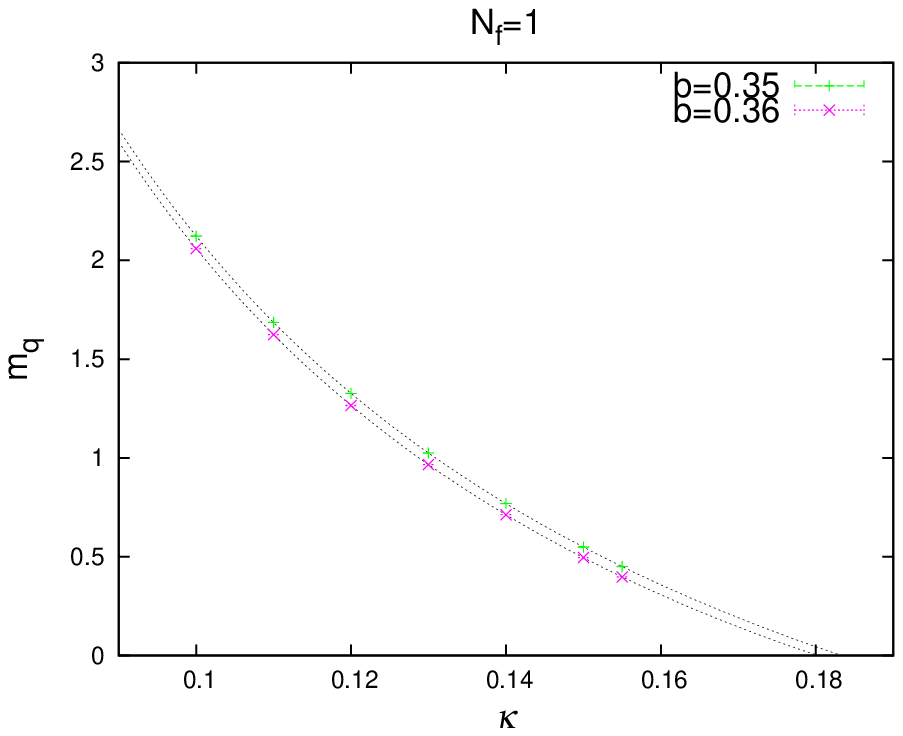}
  \end{center}
\vspace{-0.2cm}  
  \caption{$\kappa$ dependence of $m_q$ for $N_f$=1.}
  \label{fig_mq_1}
 \end{minipage}
\end{figure}

One of the simplest  quantities that one can study is the low-lying spectrum of the
square of the  hermitian Wilson Dirac matrix $Q^2=(D_W \gamma_5)^2$. The
lowest eigenvalue provides a possible definition of the quark mass 
as follows $m_q = \sqrt{\lambda} $, where we use the lattice units $a$=1.  There is a small correction here
since the boundary conditions prohibit zero-momentum states. Thus, the
lowest eigenvalue contains a small $1/N$ correction to the mass, which we
have neglected. 
On the other hand, the bare quark mass is
given by  $M_q^{(0)}=\frac{1}{2}(\frac{1}{\kappa}-\frac{1}{1/8})$.
Renormalization implies the necessity of an additive renormalization
and multiplicative renormalization of the mass. Thus, we can
parametrize the dependence as follows
\vspace{-0.1cm}
\begin{equation}
m_q= A \left( \frac{1}{2 \kappa} - \frac{1}{2 \kappa_c}\right) ^\delta
\left[ 1 + B\left( {1\over \kappa} - {1\over \kappa_c}\right) \right]
\label{fitfunc_mq}
\end{equation}
where we have included a possible $O(m_q)$ correction since we are
dealing with Wilson fermions.
For a QCD-like theory one expects $\delta=1$~\cite{DDGLPT}. 
However, if the theory has an infrared fixed-point at $\kappa=\kappa_c$,
the exponent could be different from 1. We have fitted our
parameterization to our data in the range $\kappa=0.10-0.15$.  The
data  at $\kappa$=0.05 is too far to neglect higher order corrections
in $m_q$.  On the other hand, we excluded  $\kappa$=0.16, since in that 
case the system might suffer from finite size effects. The results,
however, do not change significantly when including this value. 
Good fits are obtained at $N_f=2$ and the fitting parameters are  
$\delta=0.914(11)$, $\kappa_c=0.1744(3)$ at  $b$=0.35, and 
$\delta=0.920(14)$, $\kappa_c=0.1722(5)$
at $b$=0.36. In both cases the $\delta=1$ value is statistically
disfavored.   We have repeated the same analysis for $N_f$=1.  
Here we use the data in the range $\kappa=0.10-0.155$.  We get at $b$=0.35,
$\delta=1.010(14)$, $\kappa_c=0.1834(5)$
and at $b$=0.36,
$\delta=1.021(11)$, $\kappa_c=0.1804(3)$,
compatible with  the naively expected behavior $\delta$=1.  Thus, we
conclude   that the quark mass $m_q$ gives evidence of a different critical
behavior  for $N_f$=2 and 1. The results for $m_q$ as a function of
$\kappa$ for both cases  are displayed  
in figs. \ref{fig_mq_2} and \ref{fig_mq_1},  together with the best
fit lines.  
 
\vspace{-0.2cm}

\vspace{-0.5cm}
   
\section{String tension}

The string tension $\sigma$ is extracted from the large $R$ behavior
of the square Creutz ratio $\chi(R,R)$ as  follows:

\vspace{-0.7cm}

\begin{eqnarray}
 \nonumber
\chi(R,T) =&& -\log{ \frac{W(R+0.5,T+0.5) W(R-0.5,T-0.5)}{ W(R+0.5,T-0.5) W(R-0.5,T+0.5)} } \\ 
&& \chi(R,R) \xrightarrow[R \to \infty]{}  \sigma + {2\eta \over R^2} + {\xi \over R^4} + \cdots .
\label{sigma}  
\end{eqnarray} 
\vspace{-0.1cm}
This method  has been used successfully for the pure gauge theory (twisted
Eguchi-Kawai model) \cite{TEK3}, where the three parameter ($\sigma$,
$\eta$ and $\xi$) formula describes the data very well. For our
adjoint fermion case, the smaller effective size $L$=17 and fewer
statistics limits  the range of $R$ values that can be fitted to
Eq.~(\ref{sigma}). This introduces strong correlations among the
parameters and a rather poor  determination of the $\kappa$ dependence
of each parameter. A better  way to proceed is to fix one of the
parameters and study the evolution of the other parameters with 
$\kappa$ and the inverse `t Hooft coupling $b$. From that point view 
the best choice is $\eta$, since it is dimensionless and its value is 
connected to universal properties of an effective bosonic string
theory, not expected to depend on $\kappa$ or $b$.

\begin{figure}[htb]
\begin{center}
\includegraphics[width=0.4\textwidth]{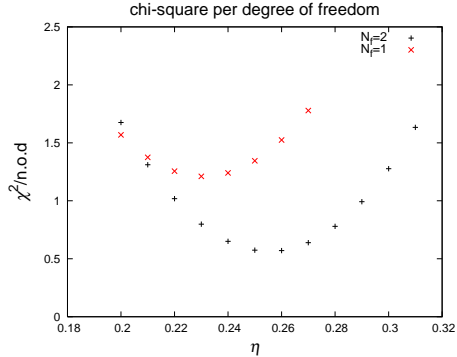}
\end{center}
\vspace{-0.7cm}
\caption{Chi-square per degree of freedom $\chi^2/n.o.d$ as functions of $\eta$ both for $N_f$=2 and 1.}
\label{fig_chi_square}
\end{figure}

To determine the value of $\eta$ to use, we perform a simultaneous fit 
to all the data (with  $\kappa$>0.05) fixing the value of $\eta$ and 
marginalizing over the  remaining parameters. 
The resulting chi-square profiles ($\chi^2/{\rm
n.o.d}$) are plotted in Fig.~\ref{fig_chi_square} for $N_f$=2 and 1.
The figure shows that our hypothesis of a common value of $\eta$ is
statistically satisfactory. The minimum of the chi-square curve
determines the best choice for $\eta$, given by $\eta$=0.26 for
$N_f$=2 and $\eta$=0.24 for $N_f$=1. The curve also provides a value
for the error of order $\pm (0.04-0.05)$.

\vspace{-0.2cm}

\vspace{-0.2cm}

\begin{figure}[htbp]
 \begin{minipage}{0.5\hsize}
  \begin{center}
   \includegraphics[width=70mm]{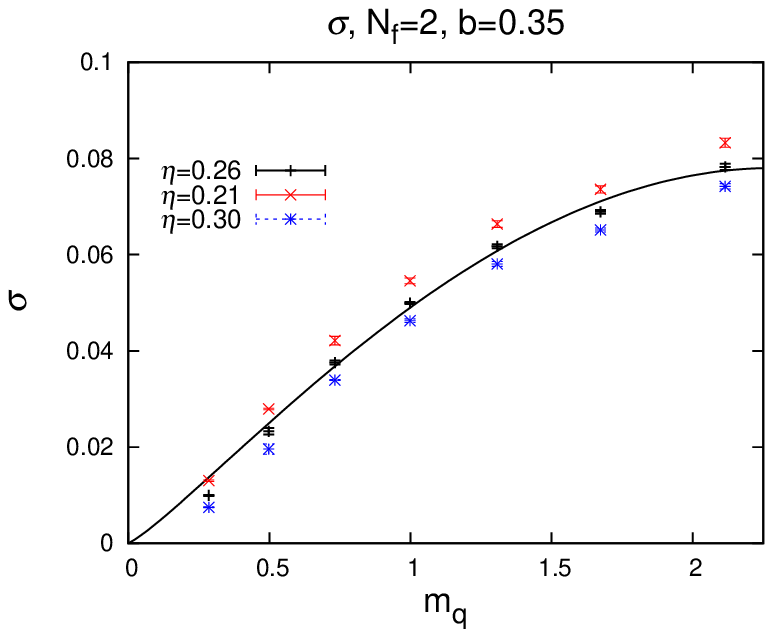}
  \end{center}
  \vspace{-0.7cm}
  \caption{$m_q$ dependence of $\sigma$ for $N_f$=2 at b=0.35.}
  \label{fig_sigma_2_35}
 \end{minipage}
 \begin{minipage}{0.5\hsize}
  \begin{center}
   \includegraphics[width=70mm]{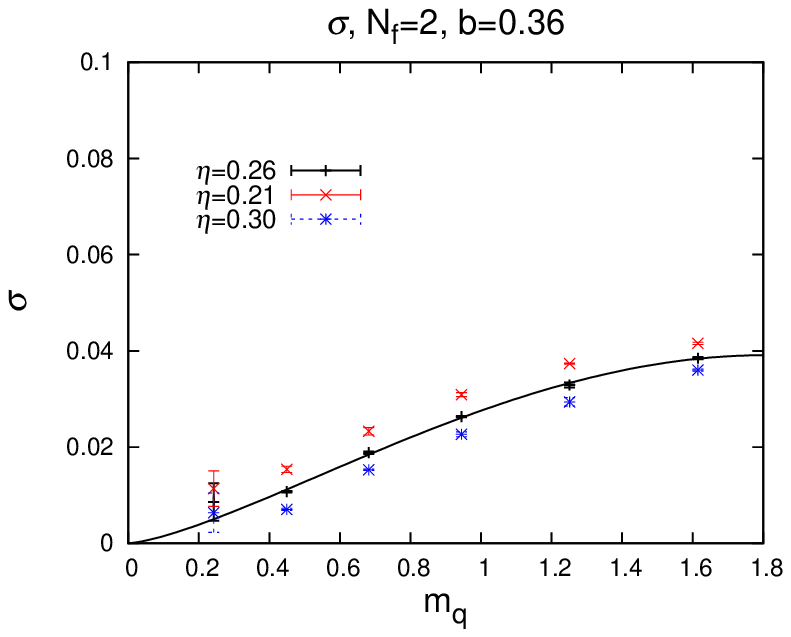}
  \end{center}
  \vspace{-0.7cm}
  \caption{$m_q$ dependence of $\sigma$ for $N_f$=2 at b=0.36.}
  \label{fig_sigma_2_36}
 \end{minipage}
\end{figure}

Fixing the value of $\eta$ we can obtain good fits to the Creutz 
ratios using the parameterization of Eq.~(\ref{sigma}). The  resulting
values of the string tension $\sigma$ as a function of $m_q$ for
$N_f$=2 at b=0.35 are displayed in Fig. \ref{fig_sigma_2_35}. The
central black symbols are obtained  with $\eta$=0.26, while red and blue symbols are 
obtained by taking $\eta$ to 0.21 and 0.30, respectively. From these
values it is clear that the string tension value depends uniformly on
$\eta$. The band covered between  the values  for $\eta$=0.21 and 0.30, 
serves as a  rough estimate of the systematic error.

If the theory is governed by an infrared fixed point deformed with a  relevant mass term 
$m_q {\bar \Psi} \Psi$, all physical quantities having positive mass
dimensions should vanish as $m_q \to 0$. In particular, the string tension 
having dimensions of mass squared should behave as  
\newpage
\vspace{-0.8cm}
\begin{equation} 
\sigma=A m_q^{\alpha}(1+B m_q)
\label{scaling_sigma}  
\end{equation}  
where we have included possible $O(m_q)$ corrections. Our $N_f=2$ data
are perfectly consistent with this formula as shown in Fig.~\ref{fig_sigma_2_35} 
for $b=0.35$ and Fig.~\ref{fig_sigma_2_36} for $b=0.36$.
Unfortunately, the exponent $\alpha$ has a large uncertainty. A fit in
the range $\kappa\in[0.10-0.15]$ for the $b=$0.35-$\eta$=0.26 data
gives $\alpha=$1.17. Varying $\eta$ within the allowed range produces
a systematic error of $0.12$. For the  $b=$0.36-$\eta$=0.26 data one
gets $\alpha=$1.42 with a systematic error of $0.25$.

\vspace{-0.2cm}

\begin{figure}[htbp]
 \begin{minipage}{0.5\hsize}
  \begin{center}
   \includegraphics[width=70mm]{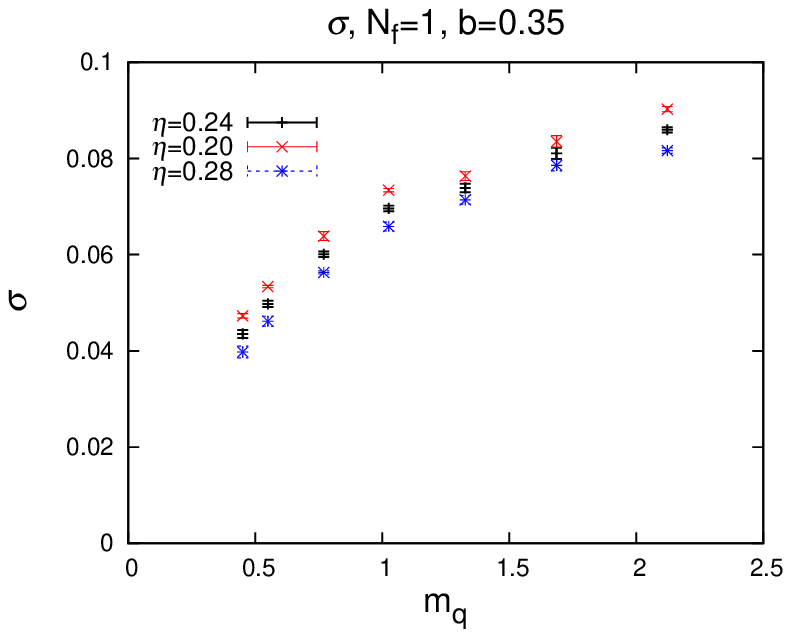}
  \end{center}
  \vspace{-0.7cm}
  \caption{$m_q$ dependence of $\sigma$ for $N_f$=1 at b=0.35.}
  \label{fig_sigma_1_35}
 \end{minipage}
 \begin{minipage}{0.5\hsize}
  \begin{center}
   \includegraphics[width=70mm]{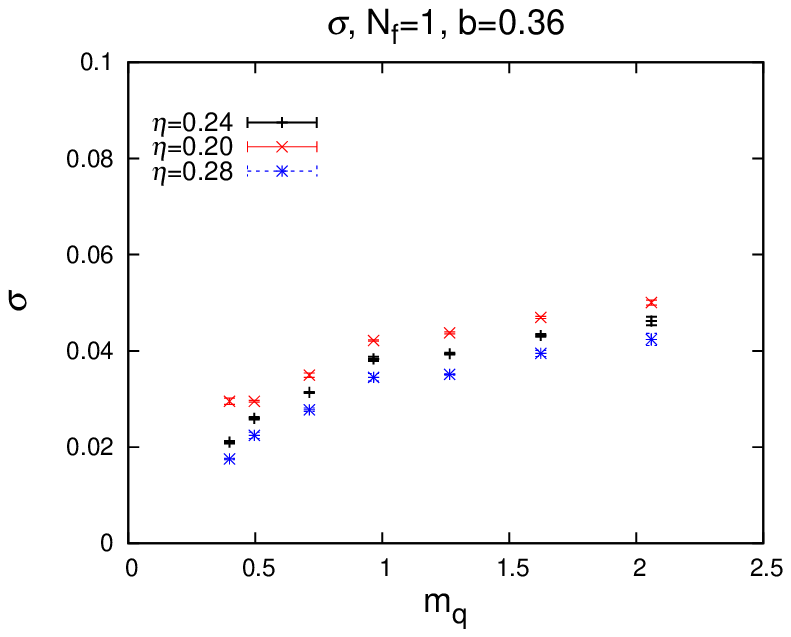}
  \end{center}
  \vspace{-0.7cm}
  \caption{$m_q$ dependence of $\sigma$ for $N_f$=1 at b=0.36.}
  \label{fig_sigma_1_36}
 \end{minipage}
\end{figure}

The previous results show that, for the $N_f=2$ case, the string tension
seems to vanish at the critical point $m_q=0$, in accordance with the 
infrared fixed point hypothesis. This contrasts with the results for
$N_f$=1, summarized in Figs.~\ref{fig_sigma_1_35} and \ref{fig_sigma_1_36}.
The data seems to approach a non-zero value at the critical point, as
expected for a QCD-like  theory, with confinement and spontaneous
symmetry breaking.

From the vanishing of the string tension at the critical point 
one can obtain a determination of the mass anomalous dimension
$\gamma_*$ at the infrared fixed point. Equating the exponent 
$\alpha'$ of $(1/\kappa-1/\kappa_c)$ with $2/(1+\gamma_*)$ one 
gets values of $\gamma_*=0.87$ and $0.53$ for  $b=$0.35 and 0.36 respectively.
The determination is dominated by the systematic error of order $\pm(0.3-0.4)$.
There is no fundamental difficulty in reducing these errors
significantly. As mentioned previously, an important source comes from
the small effective lattice volume $L^4=17^4$ of our data. However,
increasing $N=L^2$ poses an important challenge within the present
computer power. A more promising approach follows by employing 
partial volume reduction~\cite{NN,GGO}. In particular, one can 
reduce the system to a $2^4$ lattice, which at large N should behave 
as living in a $(2\sqrt{N})^4$ box. A finer analysis of the $\kappa$ 
dependence close to $\kappa_c$ is also important. Alternatively, one 
can use other observables to determine $\gamma_*$. One possibility is
to use the distribution of eigenvalues of $Q^2$. Some results 
using this method have already been presented~\cite{GGLO}.

\vspace{0.1cm}
\noindent
{\bf Acknowledgments}

A.G-A is supported from Spanish grants from MINECO: FPA2012-31686,
FPA2012-31880, FPA2009-09017, SEV-2012-0249 and CPAN CSD2007-00042;
Comunidad de Madrid: HEPHACOS S2009/ESP-1473, and European Union
PITN-GA-2009-238353 (ITN STRONGnet). M. O. is supported by 
the Japanese MEXT grant No 23540310.

The calculation has been done on Hitachi SR16000-M1 computer at High Energy Accelerator
Research Organization (KEK) supported by the Large Scale Simulation Program No.12/13-01
(FY2012-13).  The authors thank the Hitachi system engineers for their help in highly optimizing the
present simulation code.

\end{document}